\documentclass[twocolumn,showpacs,preprintnumbers,prb]{revtex4}
\usepackage{graphicx}
\usepackage{times}
\usepackage{epsfig}
\usepackage{color}
\usepackage{ulem}   %For strikethrought/strikeouts sout
\begin{document}

\def\salto{\vskip 1cm} \def\lag{\langle} \def\rag{\rangle}

\newcommand{\redit}[1]{\textcolor{red}{#1}}
\newcommand{\blueit}[1]{\textcolor{blue}{#1}}
\newcommand{\magit}[1]{\textcolor{magenta}{#1}}

\title{Systematic reduction of sign errors in many-body problems:
generalization of self-healing diffusion Monte Carlo to excited states}

\author{Fernando Agust\'{i}n Reboredo}       
\affiliation {Materials Science and Technology Division, Oak Ridge
National Laboratory, Oak Ridge, TN 37831, USA}
\begin{abstract}
  A recently developed self-healing diffusion Monte Carlo algorithm
  [PRB {\bf 79}, 195117] is extended to the calculation of excited
  states.  The formalism is based on an excited-state fixed-node
  approximation and the mixed estimator of the excited-state
  probability density.  The fixed-node ground state wave-functions of
  inequivalent nodal pockets are found simultaneously using a recursive
  approach. The decay of the wave-function into lower energy states is
  prevented using two methods: i) The projection of the improved
  trial-wave function into previously calculated eigenstates is
  removed; and ii) the reference energy for each nodal pocket is adjusted
  in order to create a kink in the global fixed-node wave-function
  which, when locally smoothed, increases the volume of the higher
  energy pockets at the expense of the lower energy ones until the
  energies of every pocket become equal. This reference energy
  method is designed to find nodal structures that are local minima
  for arbitrary fluctuations of the nodes within a given nodal
  topology.  It is demonstrated in a model system that the algorithm
  converges to many-body eigenstates in bosonic and fermionic
  cases.
\end{abstract}
\today

\maketitle
\section{Introduction}
Although several important chemical and physical properties of matter
are determined by the lowest energy electronic configuration (or
ground state), a significant number of physical properties are
crucially dependent on the excitation spectra. These properties range
from electronic optical excitations to transport and thermodynamic
behavior.

While elegant theories that take advantage of the variational
principle have been formulated for the ground state,~\cite{hohenberg,kohn} the theories on the excitation spectra are far
more complex.~\cite{hedin65} Therefore, although excited states are
extremely important, our understanding of them is limited as compared
with the ground state.

Diffusion quantum Monte Carlo (DMC) is the method of choice to obtain
the ground state energy of systems with more than $\sim\!20$
electrons. The DMC algorithm~\cite{ceperley80} transforms the
calculation of an excited state (e.g., the fermionic ground state) into
a ground state calculation. The accuracy of the method depends,
however, on a previous estimate of the zeros (nodes) of the wave-function.
 
The ground state wave-function of most many-body Hamiltonians $\mathcal{H({\bf R})}$ is a
bosonic (symmetric) wave-function without nodes.  Any other eigenstate
of a many-body Hamiltonian $\mathcal{H({\bf R})}$ must have nodes in order to be
orthogonal to the bosonic ground state. In the case of fermions 
(e.g., electrons), the ground state must be antisymmetric.  Therefore,
the electronic ground state is an excited state of the many-body
Hamiltonian $\mathcal{H({\bf R})}$ and must have nodes (hyper-surfaces in $3N_e$ space where
the wave-function becomes zero and changes sign, being $N_e$ the number of particles).

The standard diffusion Monte Carlo (DMC) approach~\cite{ceperley80}
finds the lowest energy $E^{DMC}_T$ of all the wave-functions that
share the nodes $S_T({\bf R})$ of a trial wave-function $\Psi_T({\bf
  R})$, where ${\bf R}$ is a point in the $3N_e$ coordinate space. 
This lowest energy wave-function is
denoted as the fixed-node ground state $\Psi_{FN}({\bf R})$.

Since ``no nodes'' is a condition easy to satisfy, the ground state
energy of a bosonic system can be found with a precision limited only
by statistical and time-step errors.  For any other eigenstate
$\Psi_n({\bf R})$, a good approximation of its nodal surface $S_n({\bf
  R})$ must be provided in order to avoid systematic errors.
Departures in $S_T({\bf R})$ from the exact nodes $S_n({\bf R})$
cause, in general, errors of the energy as compared with the exact
eigenstate energy.~\cite{foulkes99} For the fermionic ground state, the standard DMC
algorithm provides only an upper bound of the ground state
energy.~\cite{anderson79,reynolds82} Moreover, if $\Psi_n({\bf R})$ is
non degenerate, any departure of $S_T({\bf R})$ from $S_n({\bf R})$
creates a kink in the fixed-node ground state.~\cite{keystone}
Accordingly, accurate many-body calculations require methods to obtain
and improve $S_T({\bf R})$.  The problem of searching the exact nodes
$S_n({\bf R})$, the surfaces in
$3N_e$ space  where the wave-function of an arbitrary
eigenstate $n$ changes sign, 
%is known as ``The Sign Problem''.
is one of the outstanding problems in condensed matter theory.
~\cite{ceperley91}

This paper is the natural conclusion of earlier work.  In Ref.
\onlinecite{rosetta} we showed that even the {\it exact}
Kohn-Sham\cite{kohn} wave-functions {\it cannot} be expected to
provide accurate nodal structures for DMC calculations.  However, we
also showed that an optimal Kohn-Sham-like nodal potential exists.
Subsequently in Ref.  \onlinecite{keystone} we demonstrated that the
nodes of the fermionic ground state wave-function can be found in an
iterative process by locally smoothing the kinks of the fixed-node
wave-function. We also showed that an effective nodal potential can be
found to obtain a compact representation of an optimized trial
wave-function with good nodes. While some details are rederived here,
reading those papers before this one is {\it highly}\cite{fn:highly} recommended.

In this paper the self-healing diffusion Monte Carlo method (SHDMC) is
extended to find the nodes, wave-functions, and energies of low-energy
eigen-states of bosonic and fermionic systems.

\section{The simple SHDMC algorithm for the ground state}
This paper describes how to extend the ``simple SHDMC algorithm'' (as
described in Section III.C of Ref. \onlinecite{keystone}) to excited
states. An extension to optimize the multi-determinant expansion, (see
Section IV in Ref. \onlinecite{keystone} ) is clearly possible and
will be explained elsewhere.

The ground state SHDMC algorithm builds upon the importance
sampling DMC method.~\cite{ceperley80}  The standard diffusion Monte
Carlo approach is based on the Ceperley-Alder\cite{ceperley80}
equation:~\cite{units}
\begin{eqnarray}
\label{eq:ceperleyalder}
\frac{\partial f({\bf R},\tau )}{\partial \tau} \!  & =&  \! 
   {\bf \nabla_R^2 } f({\bf R},\tau )- \! {\bf \nabla_R }
   \left(
     f({\bf R},\tau ) {\bf \nabla_R } 
     ln
     \left|
       \Psi_T({\bf R})
     \right|^2 
   \right)
 \nonumber \\
 &&- 
 \left[
   E_L({\bf R})-E_{T}
 \right]  f({\bf R},\tau ) \; ,
\end{eqnarray}
where $E_L({\bf R}) = [\hat\mathcal{H} \Psi_T({\bf R})]/\Psi_T({\bf R}) $
is the ``local energy,'' $\hat\mathcal{H}$ is the many-body
Hamiltonian operator, ${\bf R} $ denotes a point in $3N_e$ space, and $E_T$ is a 
reference energy. 
%changes start here %FAR
Equation (\ref{eq:ceperleyalder}) is often solved
numerically\cite{ceperley80} using a large number $N_c$ of electron
configurations (or walkers) which are points ${{\bf R}_i}$ in the $3N_e$
space.  These walkers i)
randomly diffuse according to the first term in Eq.
(\ref{eq:ceperleyalder}) and ii) drift according to the second term a
time $\delta \tau$. In addition, iii) the walkers branch \{or pass on\} with
probability $p=1-\exp[(E_L({\bf R})-E_{T})\delta \tau]$  \{or $p=\exp[(E_L({\bf
  R}_i)-E_{T})\delta \tau]-1$ \}. To prevent large fluctuations
in the population of walkers and excessive branching or killing, often
a statistical weight is assigned to each walker. A
detailed review of the numerical methods used for minimizing errors
and accelerating DMC calculations is given in Ref.
\onlinecite{umrigar93}.
% FAR end changes
 
In the limit of $\tau \rightarrow \infty$, the distribution
function of the walkers in an importance sampling DMC algorithm is
given by\cite{ceperley80}
\begin{eqnarray}
\label{eq:fr}
f({\bf R},\tau \rightarrow \infty) & = & \Psi_T^*({\bf R})\Psi_{FN}({\bf R}) \; 
e^{-(E^{DMC}_T-E_T)\tau} % Added FAR
\\
           & = & \lim_{N_c \rightarrow \infty} \lim_{j \rightarrow \infty} \frac{1}{N_c} \sum_i^{N_c} 
W_i^j(j) \;
\delta \left({\bf R-R}_i^j \right) \;  . \nonumber 
\end{eqnarray}
The ${\bf R}_i^j$ in Eq. (\ref{eq:fr}) correspond to the positions of
walker $i$ at the step $j$ for an equilibrated DMC run of $N_c$
configurations. The original SHDMC method for the ground state was
implemented in a mixed branching with weights scheme. For reasons
that will be clear below, it is easier to formulate a method for
excited states with a constant number of walkers with weights
$W_i^j(k)$ which are given by
\begin{equation}
\label{eq:weights}
W_i^j(k)=e^{-\left[
E_i^j(k)-E_{T}
\right] k \; \delta\tau},
\end{equation}
with $k $ being a number of steps, $\delta\tau$ the time step, and
\begin{equation}
\label{eq:enaverage}
  E_i^j(k) = \frac{1}{k}\sum_{\ell=0}^{k-1} E_L({\bf R}_i^{j-\ell}) \; .
\end{equation} 
The energy reference $E_T$ in Eq. (\ref{eq:weights}) is adjusted so that
$\sum_i W_i^j(k)\approx N_c$ assuming a constant $E_T$ for $k$ steps.

Note that setting all $W_i^j(k)= 1$ in Eq. (\ref{eq:fr}) gives 
at equilibrium, %added %FAR
by construction, a distribution $f({\bf R})= |\Psi_T({\bf R})|^2$,
because this is equivalent to setting $E_L({\bf R})= E_{T}$ in Eq.
(\ref{eq:ceperleyalder}). 
%%Changes start here %FAR
If one sets the initial distribution of walkers as 
$f({\bf R},0)= |\Psi_T({\bf R})|^2$, then the distribution of walkers at imaginary time 
$\tau = k \delta \tau$ is given by
\begin{eqnarray}
\label{eq:evoltau}
f({\bf R},\tau) &= &\Psi_T({\bf R}) \left[ e^{-\tau \hat \mathcal{H}_{FN} } \Psi_T({\bf R}) \right] \\ \nonumber
                & = & \Psi_T({\bf R}) \Psi_T({\bf R},\tau) \\ \nonumber 
                & = & \lim_{N_c \rightarrow \infty} \frac{1}{N_c} \sum_i^{N_c} 
W_i^j(k) \delta \left({\bf R-R}_i^j \right) \; . 
\end{eqnarray}
Therefore, at equilibrium and in a no branching approach, the weights $W_i^j(k) $
contain all the difference between $f({\bf R},\tau)$ and
$|\Psi_T({\bf R})|^2$ .  In Eq.  (\ref{eq:evoltau}) 
$e^{-\tau \hat  \mathcal{H}_{FN} }$ is the fixed-node evolution operator, which is a
function of the fixed-node Hamiltonian operator $ \hat \mathcal{H}_{FN} $
given by
\begin{equation}
\label{eq:hfn}
\hat \mathcal{H}_{FN} = 
   \hat \mathcal{H}-E_T + \! \infty \ \lim_{\epsilon \rightarrow 0}
    \theta\left\{\epsilon- d_m[S_T({\bf R'})- {\bf R}] \right\} \; .
\end{equation}
The third term in the right-hand side of Eq. (\ref{eq:hfn}) adds an infinite potential 
at the points ${\bf R}$ with minimum distance to any point of the nodal surface 
$d_m[S_T({\bf R'})- {\bf R}]$  smaller than $\epsilon$.~\cite{fn:nodelta}

Using Eq. (\ref{eq:evoltau}) one can formally obtain
\begin{equation}
\label{eq:phiTtau}
\langle{\bf R}|\Psi_T(\tau)\rangle =
\Psi_T({\bf R},\tau)=
 e^{-\tau \hat \mathcal{H}_{FN} } \Psi_T({\bf R}) 
= \frac{f({\bf R},\tau )}{\Psi_T({\bf R})} \; , 
\end{equation}
and using Eq. (\ref{eq:fr}) one obtains
\begin{equation}
\langle{\bf R}|\Psi_{FN} \rangle=\Psi_{FN}({\bf R})=
\lim_{\tau \rightarrow \infty} \Psi_T({\bf R},\tau) e^{(E^{DMC}_T-E_T)\tau} \; .
\end{equation}
%Changes end here %FAR

The trial wave-function $\Psi_T({\bf R})$ 
%in Eqs. (\ref{eq:ceperleyalder}) and (\ref{eq:fr}) % FAR removed in v_5
is  commonly a product of an
antisymmetric function $\Phi_T({\bf R})$ and a
Jastrow\cite{fn:Jastrow} factor $e^{J({\bf R})}$.  Often $\Phi_T({\bf
  R})$ is a truncated sum of Slater determinants or pfaffians
$\Phi_n({\bf R})$:
\begin{equation}
\label{eq:psit}
\langle{\bf R}|\Psi_T\rangle= \Psi_T({\bf R})=e^{J({\bf R})} \sum_n^{\sim} \lambda_n
\Phi_n({\bf R})
 \; .
\end{equation}

In Ref. \onlinecite{keystone} we proved that we can evaluate 
$ e^{-\tau  \hat\mathcal{H}} |\Psi_T\rangle$ for $ \tau \rightarrow \infty$ using a
numerically stable algorithm.  
% For later use, let me express our algorithm symbolically as
The analytical derivation of the algorithm\cite{keystone} can be summarized\cite{fn:highly} here as
\begin{eqnarray} 
\nonumber
|\Psi_0\rangle & = & \lim_{\tau \rightarrow \infty} e^{-\tau \hat\mathcal{H}} 
         |\Psi_T^{\ell=0}\rangle \\
\label{eq:prevground}
         & = &
\lim_{\stackrel{\ell \rightarrow \infty}{\tau \rightarrow \infty}}  
\prod_{\ell} (e^{-\delta \tau^{\prime} \hat \mathcal{H}} 
          e^{-\tau \hat \mathcal{H}^{(\ell-1)}_{FN}}) | \Psi_T^{\ell=0}\rangle \\ % \nonumber
\label{eq:algorithmground}
         & = & 
\lim_{\stackrel{\ell \rightarrow \infty}{\tau \rightarrow \infty}}  
\prod_{\ell} (\tilde D e^{-\tau \hat \mathcal{H}^{(\ell-1)}_{FN}}) | \Psi_T^{\ell=0}\rangle \\ 
         \nonumber
         & = & | \Psi_T^{\ell \rightarrow \infty }\rangle \;.
\end{eqnarray}
The operator $\tilde D$ is defined in Eq. (\ref{eq:deltaexp}).
Equation (\ref{eq:algorithmground}) means that the ground state
$|\Psi_0\rangle $ 
%footnote here %FAR
\cite{fn:groundstate}
 can be obtained recursively by generating
a new trial wave-function $|\Psi_T^{\ell} \rangle$ from a fixed-node
DMC calculation that uses the previous trial wave-function
$|\Psi_T^{\ell-1} \rangle$, which is given by
\begin{eqnarray}
\label{eq:psitnew}
|\Psi_{T}^{\ell} \rangle & = & \tilde D
\lim_{\tau \rightarrow \infty} e^{-\tau \mathcal{H}^{(\ell-1)}_{FN}} |\Psi_T^{\ell-1} \rangle 
\\ \nonumber
& = &  \tilde D |\Psi_{FN}^{\ell} \rangle \; .
\end{eqnarray}
Equation (\ref{eq:psitnew}) means that new coefficients $\lambda_n$ of
a truncated expansion of a trial wave-function of the form given in Eq. 
(\ref{eq:psit}) are obtained {\it numerically} 
%FAR added in v_5 
from the distribution of walkers of a DMC run as
\begin{equation}
\label{eq:sampcoeff}
\langle \lambda_n \rangle = \frac{1}{N_c} \sum_{i=1}^{N_c}
W_i^j(k \gg 1) \;
\xi_n^*({\bf R}_i^j) \;
\gamma({\bf R}_i^j) \; ,
\end{equation}
where
\begin{equation}
\label{eq:xi}
\xi_n({\bf R})= e^{-2J({\bf R})} \frac {\Phi_n ({\bf R})} { \Phi_T ({\bf R})} 
\end{equation}
and \cite{keystone,umrigar93}
\begin{equation}
\label{eq:gamma}
 \gamma ({\bf R})= \frac{-1 + \sqrt{1 + 2 |{\bf v}|^2 \tau}}
{|{\bf v} |^2  \tau} 
\text{ with }
{\bf v} = \frac{\nabla \Psi_T({\bf R})} {\Psi_T({\bf R})} \;.
\end{equation}

A complete explanation of our method is given in Ref.
\onlinecite{keystone}.  Briefly here, our method systematically
improves the nodes for three main reasons:

1) The projectors in Eq. 
(\ref{eq:xi}) include only functions $\Phi_{n}({\bf R})$ that retain all symmetries of the
ground state. In more technical terms, the ground state is expanded only
with functions that belong to the same irreducible representation.  This
means that if the $\Phi_n({\bf R})$ are determinants, for example, the
bosonic ground state is excluded.  Therefore,
fluctuations that depart from the fermionic Hilbert space are filtered
and do not propagate into the trial wave-function from one DMC run
to the next  SHDMC iteration.  

2) The projection of $\Psi_{FN}({\bf R})$ into a finite set of
$\Phi_{n}({\bf R})$ with low non-interacting energy can be
shown\cite{keystone} to be equivalent to locally smoothing the kinks
at the node of the fixed-node wave-function with a function of the
form
\begin{equation}
\label{eq:deltaexp}
\langle{\bf R}|\tilde D |{\bf R^\prime}\rangle = \tilde \delta
\left({\bf R,R^\prime} \right) =
\sum_n^{\sim}
\Phi_n({\bf R}) \Phi_n^*({\bf R^{\prime}}) \; .
\end{equation}
We proved that a large class of local smoothing functions
have the same effect on the nodes as a Gaussian, under certain
conditions, which includes the
case of Eq. (\ref{eq:deltaexp}).
In turn, in Ref. \onlinecite{keystone} we proved that, to linear order in
$\sqrt{\delta\tau^{\prime}}$, the convolution of a Gaussian with any continuous function 
has the same effect on the nodes as the imaginary time propagator 
$ e^{-\delta \tau^{\prime} \hat{\mathcal{H}}}$ 
[this allows replacing Eq. (\ref{eq:prevground}) by Eq. (\ref{eq:algorithmground})]. 

Thus our method can be viewed as the recursive application of two
operators on the trial wave-function: i) $e^{-\tau \mathcal{H}_{FN}}$
that turns $|\Psi_{T} \rangle$ into $|\Psi_{FN} \rangle$ and ii)
$\tilde D$ that samples and truncates the expansion and changes the
nodes as $ e^{-\tau \hat{\mathcal{H}}}$. Accordingly, our method is
formally related to the shadow wave-function~\cite{shadow} and the
A-function approach~\cite{bianchi93,bianchi96} [see Eq. (\ref{eq:prevground})].

3) Finally, we argued that the method is robust against statistical
noise, because the kink should increase with the distance between the
exact node $S({\bf R})$ and the node of the trial wave-function
$S_T({\bf R})$ [the kink must disappear for $S_T({\bf R})= S({\bf R})$].
In addition, we took the relative error in $\lambda_n$ as truncation
criterion for $\tilde D$.

\section{Extension of the Self-Healing DMC algorithm   to excited states} 

A detailed explanation of the advantages and limitations of the
standard fixed-node approximation for excited states is given in
Ref.~\onlinecite{foulkes99} This paper explores the possibility of overcoming
these limitations in calculating excited states by excluding the
projection of lower energy states from the set of $\xi_n({\bf R})$.
However, in to follow this path the problem of inequivalent
nodal pockets has to be addressed.

\subsection{Inequivalent nodal pockets}
The expression ``nodal pocket'' denotes a volume in $3N_e$ space
enclosed by the nodal surface $S_T({\bf R})$.  It has been
shown~\cite{ceperley91} that the ground state of any fermionic
Hamiltonian with a local potential has nodal pockets that belong to
the same class, meaning that the complete $3N_e$ space can be covered by
applying all symmetry operations (e.g., particle permutations) to just
one nodal pocket.  Therefore, if the trial wave-function is obtained
from such a Hamiltonian, all nodal pockets are equivalent by symmetry.
For the ground state, one can obtain the fixed-node wave-function in
just one pocket and map it to the rest of the $3N_e$ space using
permutations of the particles and other symmetries of
$\hat\mathcal{H}$.

In the case of arbitrary excited states, there are inequivalent nodal
pockets that present a challenge to the fixed-node
approach.~\cite{HLRbook}  Due to this inequivalent pocket problem,
alternatives to the fixed-node method and variations have been
tried.~\cite{ceperley88,barnett91,blume97,dasilva01,nightingale00,luchow03,
  schautz04,umrigar07,purwanto09}  Self-healing DMC\cite{keystone}
implicitly takes advantage of the equivalence of nodal pockets in the
fermionic ground state and must be extended to the inequivalent pocket
case. For this reason a nonbranching formulation is used in the
excited state case.

\subsection{Equilibration of walkers in inequivalent nodal pockets}
A first complication, which has a simple solution, of the
nonbranching fixed-node approximation is that the number of walkers
in each nodal pocket is also fixed by the nodes. As a consequence of
the drift or ``quantum force'' term [second term in Eq.
(\ref{eq:ceperleyalder})], the walkers are repelled from the regions
where the wave-function is zero and they cannot cross the node for
$\delta \tau \rightarrow 0$. The fact that the population in each
nodal pocket is fixed has no consequence for the ground state because
all nodal pockets are equivalent. For the ground state it is not
important in which nodal pocket the walker is trapped because particle
permutations can move every walker into the same nodal pocket and the
projectors $\xi_n({\bf R})$ in Eq. (\ref{eq:xi}) are invariant under
such permutations.

However, in the case of excited states, which have more nodes than
those required by symmetry,~\cite{fn:permutations} there are
inequivalent nodal pockets. In a nonbranching DMC scheme with
weights, the population is locked from the start in a set of pockets.
If the initial distribution of $N_c$ walkers is chosen with a
Metropolis algorithm to match $|\Psi_T({\bf R})|^2$, there would be
random variations in the starting population of the order of
$\sqrt{N_c/N_p}$, where $N_p$ is the number of inequivalent nodal
pockets.  This would cause systematic errors if the wave-function
coefficients $\lambda_n$ were sampled without taking preventive
measures.  Moreover, even if the initial numbers of walkers in each
pocket were set ``by hand'' (to be proportional to the integral
$|\Psi_T({\bf R})|^2$ in each pocket), the resolution of the sampling
cannot be better than $1/N_c$.  The importance of this error grows if
$N_c$ is small or if the number of inequivalent nodal pockets is
large.

To prevent this error from occurring, some walkers are simply
allowed to cross the node after the wave-function coefficients are
sampled. At the end of a sub-block of $k$ steps, for every walker $i$
at ${\bf R}_i$, a random move ${\bf \Delta R}_i$ is generated with a
Gaussian distribution using $\sigma^2= \delta\tau^{\prime}$, {\it
  without} the drift velocity contribution.  This move is accepted
only if the wave function changes sign with a Metropolis probability
$p=\max\left\{ 1,[\Psi_T({\bf R}_i {\bf +\Delta R}) /\Psi_T({\bf
    R}_i)] ^2 \right\} $. This ensures that i) the distribution 
of walkers remains proportional to $|\Psi_T({\bf R})|^2$ and ii) the
average number of walkers in each pocket is proportional to the
integral of $|\Psi_T({\bf R})|^2$ as the number of sub-blocks $M$
tends to $\infty$.
 
\subsection{Unequal fixed-node energies in inequivalent nodal pockets}

 A second complication of the fixed-node approach 
for the general case of excited states appears because small
departures of $S_T({\bf R})$ from the exact nodes $S_n({\bf R})$ often
will result in inequivalent nodal pockets having fixed-node solutions with
different fixed-node energies. When nodal pockets are not equivalent, a
standard DMC algorithm will converge to a ``single nodal pocket''
population.  In this case, the lowest energy pocket will contain
all the walkers in a branching algorithm [or all significant weights
($W_i^j(k) \ne 0 $ )].  Accordingly, the average energy sampled will
correspond to the lowest energy nodal pocket, which will be different
from that of the true excited-state energy (see Chapter 6 
in Ref. \onlinecite{HLRbook} and references therein). 

If the coefficients of an excited-state fixed-node wave-function are
sampled with the same procedure used for the ground
state\cite{keystone} [see Eq. (\ref{eq:sampcoeff})], they would
correspond to a function that is different from zero just at the class
of nodal pockets with lowest DMC energy and zero everywhere else. This
function will not be, in general, orthogonal to the lower energy
states. Moreover, this will result in kinks at the nodes in the
wave-function sampled with Eq. (\ref{eq:sampcoeff}) between lowest
energy nodal pockets and inequivalent ones.

A first preventive measure to avoid a single pocket population is to
avoid 
%propagation to infinite imaginary time. 
%FAR added in v_5
the limit $\tau \rightarrow \infty$ in
Eqs. (\ref{eq:algorithmground}) and (\ref{eq:psitnew}) which 
replaces $|\Psi_{FN}^{\ell} \rangle $ by 
$e^{-k \delta \tau \mathcal{H}^{(\ell-1)}_{FN}} |\Psi_T^{\ell-1} \rangle $ 
in Eq. (\ref{eq:psitnew}).
As a result $k$ in Eq. (\ref{eq:sampcoeff}) 
is limited to small values, which
brings all values of $W_i^j(k)$ closer to $1$.
% end added FAR v_5
Since the approach is
recursive, the limit of $\tau \rightarrow \infty$ is reached as 
$\ell \rightarrow \infty$ (since
successive applications of the algorithm are accumulated in
$|\Psi_T^{\ell}\rangle$). In addition, to prevent the wave-function from falling
into lower energy states, two techniques are used: i) direct
projection and ii) unequal reference energies.

\subsection{Direct projection} 
While the trial wave-function can be forced to be orthogonal to the
ground state, or any other excited state calculated before, the
fixed-node wave-function can develop a projection into lower energy
states, because the DMC algorithm only requires $\Psi_{FN}({\bf R})$
to be zero at the nodes $S_{T}({\bf R})$. To prevent excited
states from drifting into lower energy states, let me assume, for a moment,
that approximated expressions of the excited states $ \langle{\bf R}|e^{\hat
  J}|\breve\Phi_n\rangle = \Psi_n({\bf R})=e^{J({\bf R})}\breve \Phi_n({\bf
  R})$ with $n \le \nu$ can be obtained and used to build the
projector
\begin{equation}
\label{eq:orthogonality}
\hat P = e^{\hat J}\left[ 1 - \sum_n^{\nu} 
|\breve \Phi_n\rangle \langle\breve \Phi_n^*|\right]   e^{-\hat J} \; \;,
\end{equation}
where the operator $e^{\hat J}$ is the multiplication by a Jastrow.
Since the $|\breve \Phi_n\rangle $ shall be obtained statistically, they
will have errors and will not form an orthogonal basis in
general.  Therefore, $\langle\breve \Phi_n^*|$ are elements of the
conjugated basis that satisfy $\langle\breve \Phi_n^*|\breve \Phi_m\rangle =
\delta_{n,m}$. They can be constructed inverting the overlap matrix
$S_{n,m}= \langle\breve \Phi_n|\breve \Phi_m\rangle$ as
\begin{equation}
\langle\breve \Phi_n^*|= \sum_m S^{-1}_{n,m} \langle\breve \Phi_m| \; .
\end{equation}
 Then, the extension of the self-healing
algorithm to the next excited $|\Psi_{\nu+1}\rangle$ can be rederived 
analytically as follows:
\begin{eqnarray}
\label{eq:SHDMCexcited}
\nonumber
|\Psi_{\nu+1}\rangle & = & 
\lim_{\tau \rightarrow \infty} \hat P \; e^{-\tau \hat\mathcal{H} }\hat P  
|\Psi_{T,\nu+1}^{\ell=0}\rangle \\
           \nonumber
         & = &  
\lim_{\ell \rightarrow \infty} 
\hat P \;
\prod_{\ell} 
\left( 
 e^{-(\delta \tau^{\prime}+k \delta\tau ) \hat \mathcal{H}} \hat P
\right) 
| \Psi_{T,\nu+1}^{\ell=0}\rangle
\\
           \nonumber
         & = &  
\lim_{\ell \rightarrow \infty} 
\hat P \;
\prod_{\ell} 
\left( 
 e^{-\delta\tau^{\prime}  \hat \mathcal{H}}  
 e^{-k \delta\tau  \hat \mathcal{H}_{FN}^{(\ell-1)}} 
\hat P
\right) 
| \Psi_{T,\nu+1}^{\ell=0}\rangle
 \\ 
         & \simeq &  
\lim_{\ell \rightarrow \infty} 
\hat P \;
\prod_{\ell} 
\left( 
 \tilde D e^{-k \delta\tau  \hat \mathcal{H}^{(\ell-1)}_{FN}} \hat P
\right) 
| \Psi_{T,\nu+1}^{\ell=0}\rangle \\ \nonumber
         & = & | \Psi_{T,\nu+1}^{\ell \rightarrow \infty }\rangle.
\end{eqnarray}
%FAR added
Equation (\ref{eq:SHDMCexcited}) means that for any initial trial
wave-function $|\Psi_{T,\nu+1}^{\ell=0}\rangle$ with $\hat P
|\Psi_{T,\nu+1}^{\ell=0}\rangle \ne 0 $, one can obtain the next excited
state $|\Psi_{\nu+1}\rangle$ recursively.  The numerical
implementation of the algorithm for excited states (see Section
\ref{sc:algorithm} for details) is almost identical to the ground state
version\cite{keystone} with three differences: i) there is no
branching and the product $k \delta \tau$ is chosen so as $W_i^j(k)
\simeq 1 $ [see Eq. (\ref{eq:sampcoeff})], ii) the projection of the
vector of coefficients $\lambda_n$ into the ones corresponding to
eigenstates calculated earlier is removed with $\hat P$, and iii) some
walkers cross the node after $k$ time steps (see above).
%FAR

Eq. (\ref{eq:SHDMCexcited}) holds in the limit of $N_c \rightarrow
\infty$, $\delta \tau \rightarrow 0$, $\delta \tau^{\prime} \rightarrow 0$, $\ell k \delta \tau
\rightarrow \infty$, and $\ell \delta \tau^{\prime}
\rightarrow \infty$.
In the derivation of Eq. (\ref{eq:SHDMCexcited}),
the following properties were used: $\hat P^2=\hat P$, and  
$[\hat \mathcal{H},\hat P] \simeq 0$.
In Ref. \onlinecite{keystone} it was shown that, under certain conditions,
\begin{equation}
S \left[
 e^{- \delta\tau^{\prime}  \hat \mathcal{H} } e^{-k \delta\tau  \hat \mathcal{H}^{(\ell-1)}_{FN}} \hat P
| \Psi_T^{\ell}\rangle 
\right ]
\simeq
S \left[
 \tilde D e^{-k \delta\tau  \hat \mathcal{H}^{(\ell-1)}_{FN}} \hat P
| \Psi_T^{\ell}\rangle
\right ] \; ;
\end{equation}
that is, the nodes of the two functions in the brackets are approximately
the  same. 

Note that the second term in brackets of Eq.
(\ref{eq:orthogonality}) has precisely the form given in Eq.
(\ref{eq:deltaexp}). By construction, this term would generate
a function with nodes corresponding to a linear combination of lower
energy eigenstates. The projector $\hat P$, instead, excludes any
change in the wave-functions introduced by the projection and sampling operator
$\tilde D$ or by $e^{-\tau \mathcal{H}^{(\ell-1)}_{FN}} $ in the
direction of lower energy wave-functions (which includes their nodes).

\subsection{Adjusting the reference energy in each nodal pocket} 
If walkers at one side of the node have more weight than at the other
(because of inequivalent pockets with different fixed-node energies),
the propagated wave-function obtained by sampling the walkers will be
multiplied by a larger (smaller) factor for the low (high) energy side
of the nodal surface.  This generates an additional contribution to
the kink at the node that, when locally smoothed, increases the
volume of lower energy pockets at the expense of the higher energy
ones, causing the volume of the lower (higher) energy pockets to grow
(diminish).  This, in turn, will have an impact on the kinetic energy:
due to quantum confinement effects, the difference in fixed-node
energies will increase in the next iteration. This very
interesting effect in fact acts to our advantage by helping us to find
the ground state even when starting from a very poor
wave-function.~\cite{keystone} For excited states, this effect is
prevented by i) limiting the maximum value of $k$ and ii) the
projector $\hat P$ in Eq.  (\ref{eq:SHDMCexcited}).  However, the
eigenstates $|\Psi_n\rangle$ will have statistical errors that can
create systematic errors in the higher states. To partially prevent
these errors, and to limit the number of orthogonality
constraints, the energy reference can be changed in order to invert
this contribution to the kink to our advantage.

While a single reference energy $E_T$ can still be used for the DMC
run in each block, the projectors of Eq. (\ref{eq:sampcoeff}) are redefined
using a reference energy  dependent on the nodal pocket.  In addition,
following a suggestion of C.  Umrigar,~\cite{umrigar_private} the
change in the coefficients $\delta \lambda_n $ is sampled instead of the
total value $ \lambda_n $.
\begin{eqnarray}
\label{eq:sampnew}
\lambda_n^{\ell} &= & \lambda_n^{\ell-1}+ \langle \delta \lambda_n \rangle \\ \nonumber
\langle \delta \lambda_n \rangle & = & 
\frac{1}{N_c} \sum_{i=1}^{N_c}
(W_i^j(k) e^{-\beta 
\left[
 E_{T}-\bar E_i^j(j_0)
\right]k \; \delta\tau
} -1)\;
\xi_n^*({\bf R}_i^j) \;
\gamma({\bf R}_i^j) \; ,
\end{eqnarray}
where $\beta$ is an adjustable parameter and
\begin{equation}
\bar E_i^j(j_0) = 
\frac{\sum_{m=j_0}^j W_i^m(k)\gamma({\bf R}_i^m) E_L({\bf R}^m_i)}
{\sum_{m=j_0}^j W_i^m(k)\gamma({\bf R}_i^m) }
\end{equation}
is the weighted average of the local energy during the lifetime of the
walker $i$ since the start of the block or the last time it crossed
the node at step $j_0$.  If $\beta = 1 $ is selected in Eq.
(\ref{eq:sampnew}), the factor $e^{-\beta [ E_{T}-\bar E_i^j(j_0) ]}$
just replaces in the definition of the weights [see Eq.
(\ref{eq:weights})] $E_{T}$ by $\bar E_i^j(j_0)$.  The energy $\bar
E_i^j(j_0)$ for $j-j_0 \gg k$ is expected to converge to the fixed-node energy
of the nodal pocket where the walker $i$ is trapped; however, only the last
two-thirds of the block are used to accumulate values to allow $\bar
E_i^j(j_0)$ to equilibrate.

It was argued before that, for $\beta = 0 $, the differences in the
fixed-node energies of neighboring nodal pockets create a
contribution to the kink that, when locally smoothed, increases the volume
of nodal pockets with low fixed-node energy. For $\beta > 1$, it is
likely that this contribution to the kink is inverted so that the volume
of the lower (higher) energy pockets is reduced (increased) by the
smoothing function (\ref{eq:deltaexp}).  Therefore, it can be assumed
that a value of $\beta > 1$ should stabilize the higher energy nodal
pockets, increasing their volume and, thus, reducing their energy.  This
process will stop when the fixed-node energy of all nodal pockets 
becomes equal.

Note that by introducing this artificial contribution to the kink, one
may stabilize some nodal structures, preventing nodal fluctuations 
that reduce the energy of one nodal pocket at the expense of the
others.  However, fluctuations that lower the energy of every nodal
pocket are not prevented. Therefore, if several eigenstates have the
same nodal topology, higher energy states could drift into lower
energy ones if orthogonality constraints [see Eq.
(\ref{eq:orthogonality})] are not imposed.

Finally, note that choosing $\beta > 1$ can also cause problems if the
quality of the wave-function is not good or if the statistics is poor.
For example, a small statistical fluctuation in the values of
$\lambda_n$ could create a new nodal pocket with high energy. In
successive blocks (as $\ell$ increases), this pocket will grow at the
expense of the others, causing the total energy to rise.

\section{SHDMC algorithm for excited states}
\label{sc:algorithm}
A basis of $\Phi_n({\bf R})$ must be constructed, taking advantage of
all the symmetries of $\hat \mathcal{H}$.\cite{fn:permutations} The
$\Phi_n({\bf R})$ should be selected to be eigen-functions of a
noninteracting many-body system~\cite{keystone} belonging to the same
irreducible representation for every symmetry group of $\hat
\mathcal{H}$.  The calculation must be repeated for each irreducible
representation.  Note that the same algorithm is used for bosons or
fermions: the only difference is the basis used to expand the
wave-functions.
  
  The calculation of excited states with SHDMC is composed of a
  sequence of blocks. Each block $\ell$ has $M$ sub-blocks with $k$ standard
  DMC steps. 
  
The basic algorithm is the following:
 
\begin{enumerate}
 \item An initial set of coefficients for the expansion of the trial wave-function is selected.
   
 \item The changes $\delta \lambda_n$ are accumulated [see Eqs.
  (\ref{eq:xi}) and (\ref{eq:sampnew})] at the end of each  sub-block.
  Some walkers near the node can cross it at the end of each
  sub-block.
  
\item At the end of each block $\ell$, the error in $\delta \lambda_n$
  is evaluated. If this error is larger than 25\% of $\lambda_n
  + \delta \lambda_n$, then $\lambda_n $ is set to zero;~\cite{keystone}
  otherwise, $\lambda_n $ is set to $\lambda_n + \delta \lambda_n$.

 \item A new trial wave-function is constructed at the end of each
  block $\ell$ using the new values of the coefficients sampled 
  %FAR added in v_5
  after
  removing with $\hat P$ the projection into eigenstates calculated earlier.  
  %FAR end add
  
\item If the scalar product between the vector of new
  $\delta\lambda_n$ with the one obtained in the previous block
  ($\ell-1$) is positive, the number of sub-blocks $M$ is increased by
  one. Otherwise, $M$ is {\it multiplied} by a factor larger than one
  (e.g., $1.25$). This factor increases the statistics reducing the
  impact of noise.~\cite{fn:changes}
  
\item Steps 2-6 are repeated until the variance of the weights
  $W_i^j(k)$ is smaller than a prescribed tolerance (see Fig.
  \ref{fg:variance} in Section \ref{sc:modeltest}).

 \item The projector $\hat P$ is updated to include the new excited state.

 \item Steps 1-7 are repeated until a desired number of excited states is obtained.
\end{enumerate}
\subsection{Remarks}
Some points about the application of the algorithm should be addressed
before discussing the results.
\begin{itemize} 
\item In this paper, to test the method, intentionally
  poor trial wave-functions have been selected as a starting point.  
  Good initial wave-functions and a good Jastrow are advised in
  real production runs in large systems. Methods to select good
  initial trial wave-functions will be discussed elsewhere. 
\item Time-step errors and, in particular, persistent walker
  configurations\cite{umrigar93} can cause significant problems.  When
  this happens it often results in an increase in the error bar of
  every $\lambda_n$ which causes a large reduction in the number of
  coefficients retained in the trial wave-function. This problem is
  avoided in the algorithm by discarding the entire block if a 50\%
  reduction in the number of basis functions retained is detected.
  Nevertheless, if the quality of the initial $\Psi_T({\bf R})$ is
  bad, it is strongly recommended to reduce the time step $\delta
  \tau$. As the quality of the wave-function improves with successive
  iterations, one can increase $\delta\tau$. For fast convergence
  $\sqrt{k \;\delta\tau}$ should be of the order of the interparticle
  distance.
\item As a strategy, it is better to run at first using $\beta=0$ in
  Eq. (\ref{eq:sampnew}) including every state calculated before in
  $\hat P$ [see Eq. (\ref{eq:orthogonality})]. Once the wave-function
  $\Psi_T({\bf R})$ is converged, one can set $\hat P=1$ and $\beta=1$
  and monitor if $\Psi_T({\bf R})$ evolves into a subset of lower
  energy states. To prevent the propagation of errors of
  every lower energy state included in $\hat P$ into the next excited
  state, a run including only this subset in $\hat P$ can be
  performed.
\item To obtain accurate total energies, a long run with large
  $k$ is required (this is almost a standard DMC run).
\item SHDMC should not be used blindly as a library routine.  The
  calculation of excited states with SHDMC is a task that will
  probably remain limited to quantum Monte Carlo experts. While, in
  contrast, density functional approximated methods have suddenly
  become very easy to use, it is not quite clear to the author that
  requiring expertise and a deep understanding is a disadvantage. Any
  new code using SHDMC should be tested in a small system where
  analytical solutions or results with an alternative
  approach\cite{umrigar07} are available.  The comparison with a
  soluble model is presented in the next section.
\end{itemize}
\section{Applications to Model Systems}
\label{sc:modeltest}
This section compares the methods
described above for the calculation of excited states with SHDMC, with full
configuration interaction (CI) calculations in the model system used
in Refs. \onlinecite{rosetta} and \onlinecite{keystone}.

Briefly, the lower energy eigenstates are found for two electrons
moving in a two dimensional square with a side length $1$ with a
repulsive interaction potential of the form\cite{units} $V({\bf
  r},{\bf r^{\prime}}) = 8 \pi^2 \gamma \cos{[\alpha
  \pi(x-x^{\prime})]}\cos{[\alpha \pi(y-y^{\prime})]}$ with $\alpha=
1/ \pi$ and $\gamma = 4$.  The many-body wave-function is expanded in
functions $\Phi_n({\bf R})$ that are eigenstates of the
noninteracting system. The $\Phi_n({\bf R})$ are linear combinations of
functions of the form $\prod_{\nu}  \sin(m_{\nu} \pi x_{\nu})$ with
$m_{\nu} \le 7$.  Full CI calculations are performed to obtain a nearly
exact expression of the lower energy states of the system $\Psi_n({\bf
  R})= \sum_m a_m^n \Phi_m({\bf R})$.

We solve the problem both for the singlet and the triplet case.  The
singlet state of this system is bosonic-like, since the ground state
wave-function has no nodes.  The lowest energy excitations of the
noninteracting problem $\Phi_n({\bf R})$ that have the same symmetry
(that is, that are invariant under exchange of particles, and under
all symmetry operations of the group $D_4$) are selected to expand
$\hat \mathcal{H}$.  For the case of the triplet, the wave-function
must change sign for permutations of the particles.  The ground state
is, however, degenerate (belongs to the $E$ representation of $D_4$).
The $E$ representation can be described by wave-function even
(odd) for reflections in $x$ and odd (even) for reflections in $y$. We
choose the wave-functions that are odd in the $x$ direction: belonging
to a $D_2$ subgroup of the $D_4$ symmetry. For more details on the
triplet ground state calculations, see Refs.  \onlinecite{rosetta} and
\onlinecite{keystone}.

To facilitate the comparison with the full CI results,
projectors $\xi_n({\bf R})$ are constructed with the same basis functions
used in the CI expansion.  For the same reason, no Jastrow
function is used [$J=0$ in Eq.~(\ref{eq:xi})].  

To test the method, poor initial trial wave-functions are intentionally
chosen as follows: For the ground state the lowest energy
function of the noninteracting system is selected.  For the $n^{th}$ ($n =
\nu+1$) excited state, the initial trial wave-function $
|\Psi_{T,n}^{\ell=0}\rangle $ was constructed by completing the first
$\nu$ columns of a determinant with the first $\nu+1$ coefficients of
the $\nu$ eigenstates calculated before.  Subsequently, the vector of
cofactors of the last column was calculated. The coefficients of this
vector are used to construct a trial wave-function orthogonal to all
the eigenstates calculated earlier.

\begin{figure}
\includegraphics[width=1.00\linewidth,clip=true]{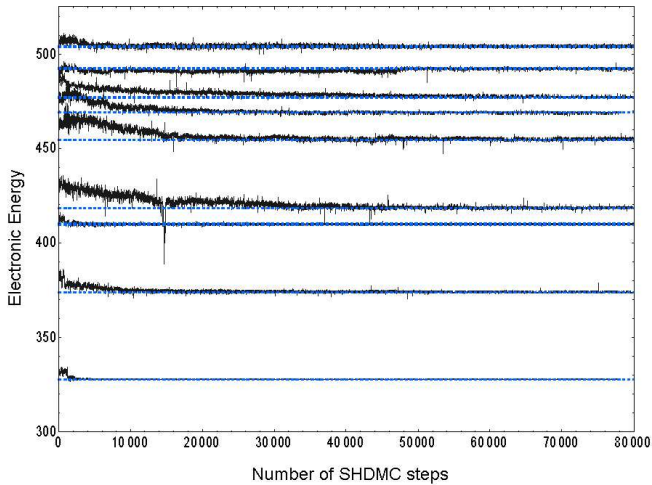}
\caption{(Color online) Self-healed DMC run obtained 
  for successive eigenstates belonging to the $A_1$ (trivial) irreducible
  representation of the group $D_4$ in the singlet state.  Black lines
  denote the average value of the local energy. The horizontal blue
  dashed lines mark the energy of the corresponding excitation in the
  full CI calculation.
\label{fg:dmcCIrun:sglex}}
\end{figure}

Figure \ref{fg:dmcCIrun:sglex} shows the results of successive SHDMC
runs for the singlet ground state and the next $8$ excitations that
belong to the same symmetry (total spin $S=0$, and irreducible
representation $A_1$ of the group $D_4$). The SHDMC calculations were
done using $N_c = 200$ walkers with a sub-block length $k=50$, a time step
$\delta \tau = 0.0002$,~\cite{units} $\delta \tau^{\prime} = 0.002$
(for the ground state $\delta \tau^{\prime} = 0$ ) and, $\beta =1
$ in Eq.  (\ref{eq:sampnew}).

The lines in Fig. \ref{fg:dmcCIrun:sglex} join the values obtained for
the weighted average of the local energy $E_L({\bf R})$ for each time
step.  The horizontal dashed lines mark the energy of the nearly
analytical result obtained with full CI.  The agreement between SHDMC
and full CI is extremely good. As higher energy eigenstates are
calculated however, and the number of nodal pockets and nodal surfaces
increases, time step errors start to play a dominant role. In
particular, for the $9^{th}$ excitation (not shown) $\delta \tau $
must be reduced.

The
occasional peaks (or drops) observable in the data are correlated with the update of
$\Psi_T({\bf R})$, and their reduction also reflects a systematic
improvement in the trial wave-function.
At the end of each block, the trial wave-function coefficients
$\lambda_n$ are updated and all weights are reset to 1. They gradually
reach equilibrium values when new energies are sampled, completing a
sub-block of length $k$. As a result, at the beginning of each block,
the energy sampled is the average of the trial wave-function energy, which
is often different than the DMC energy sampled thereafter (but it 
can be smaller or higher for a bad trial wave-function with small $N_c$). 

One interesting result is that some orthogonality constraints are not
required to obtain some excited states. This is the case, for example,
of the first excited state calculated with $\beta = 1 $. This is
presumably due to the fact that the number of nodal pockets is different
for the excited state and the ground state and the decay path from the first excited state
to the ground state is obstructed by the formation of a kink between
inequivalent nodal pockets if a value of $\beta \approx 1 $ is used. This is
also the case for states $6$ and $7$, which were obtained {\it before}
state 5 despite the fact that they have higher energy. 

A similar effect is observed in some triplet excitations. Due to the
choice of initial trial wave-function and the kink induced by
$\beta=1$, the $3^{rd}$ excitation is found before the $2^{nd}$, and the
$5^{th}$ is obtained before the $2^{nd}$ and the $4^{th}$. This interesting
effect disappears if $\beta=0$ is chosen.

Table \ref{tb:overlap} shows the logarithm of the residual
projection 
\begin{equation}
\label{eq:lrp}
L_{rp}=\log\left(1-|\langle\Psi_n^{CI}|\Psi_n\rangle| \right) 
\end{equation}
of the excited state wave-function $|\Psi_n\rangle$ sampled with SHDMC onto
the corresponding full CI result $|\Psi_n^{CI}\rangle$ as a function of the
number of iterations for different eigenstates. The states are ordered
as they first appear in the calculation. 

% begin add v_6 FAR
In addition, Table \ref{tb:overlap}  compares the 
values of the eigen-energies obtained with CI and SHDMC. The agreement is 
very good. In some cases the difference is larger than the error bar. This might
signal that small nodal errors remain. Note that there is no upper bound theorem
for excited states but for the ground state within an abelian irreducible 
representation.\cite{foulkes99}
% end add v_6 
\begin{table}
\caption{Values obtained for $L_{rp}$ [see Eq. (\ref{eq:lrp}) ] for 
a total of 
(a) $4 \times 10^4$  (b) $8 \times 10^4$ and (c)  $12 \times 10^4$  DMC steps 
and  corresponding eigen-energies  for two electrons in a square box with a model interaction. 
The logarithm of the residual projection $L_{rp}$  
of the SHDMC wave-function with the corresponding full result CI
is given
for different eigenstates belonging to the same symmetry of the ground
state as a function of the number of steps used to sample 
the wave-function. \label{tb:overlap} The states are included in the order they
were obtained. }
\begin{tabular}{|c|c|c|c|c|c|c|ll|}
\hline
State & Spin &Rep.  &$L_{rp}$ &$L_{rp}$         &$L_{rp}$    & CI & SHDMC &  \\
      &      &       &   a    &   b   &    c       & Energy  & Energy & \\
\hline
0 & S &A$_1$ &-14.84  & -15.05 &       &  328.088  & 328.089 &(2) \\
1 & S &A$_1$ & -6.80  & -8.85  &       &  374.106  & 374.103 &(6)\\
2 & S &A$_1$ & -7.23  & -8.69  &       &  409.960  & 409.954 &(3)\\
3 & S &A$_1$ & -4.42  & -6.07  &       &  418.508  & 418.66  &(2)\\
4 & S &A$_1$ & -3.65  & -5.01  &       &  454.630  & 454.84  &(2)\\
6 & S &A$_1$ & -.--   & -4.85  & -6.22 &  477.019  & 477.100 &(5)\\
7 & S &A$_1$ & -3.90  & -5.26  &       &  492.216  & 491.98  &(1)\\
5 & S &A$_1$ & -5.60  & -6.17  &       &  468.854  & 468.845 &(13)\\
8 & S &A$_1$ & -5.09  & -6.49  &       &  503.805  & 503.92  &(1)\\
\hline \hline
0 & T &E     & -8.49  & -8.71  &       &  342.137 & 342.191 &(5)\\
1 & T &E     & -4.37  & -4.35  &       &  385.908 & 387.80  &(1)\\
3 & T &E     & -3.06  & -3.35  &       &  422.670 & 423.60  &(2)\\
5 & T &E     & -4.04  & -5.48  &       &  438.791 & 438.70  &(1) \\
2 & T &E     & -2.31  & -2.31  &       &  411.887 & 416.07  &(1)\\
\hline
\end{tabular}
\end{table}

\begin{figure}
\includegraphics[width=1.\linewidth,clip=true]{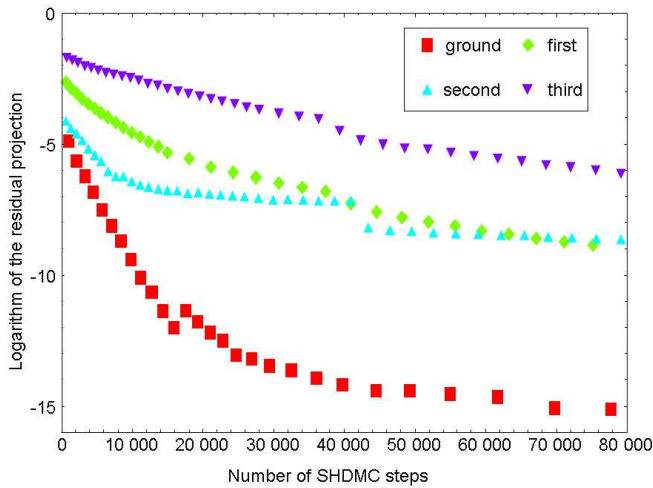}
\caption{(Color Online) Logarithm of the residual projection [see Eq. (\ref{eq:lrp})]
  for the ground (square), first (diamond), second (up triangle) and 
  third (down triangle) eigenstates with $A_1$
  symmetry and S=0. \label{fg:lrp} }
\end{figure}
Figure \ref{fg:lrp} shows $L_{rp} $ at the end of each block
for the ground state and low-lying excitations of the system as a
function of the total number of SHDMC steps. The calculations were done by first 
running $\sim\!40\,000$ SHDMC steps for each eigenstate before 
starting the calculation of the next. Subsequently, an additional
set of $\sim\!40\,000$ SHDMC steps was run, improving the projector $\hat P$.
The kinks in the data around
$\sim~40\,000$ are due to the changes in the coefficients of the lower
energy states involved in $\hat P$ [see Eq. (\ref{eq:orthogonality})].

One important conclusion of Table \ref{tb:overlap} and Figure
\ref{fg:lrp} is that errors in the determination of lower energy
states calculated earlier  only propagate ``locally'' because of the
orthogonality constraints in Eq.  (\ref{eq:orthogonality}). This error
does not have a strong impact on much higher energy excitations. This
is apparently due to the fact that each newly calculated excitation
tends to occupy the Hilbert space left by lower excitations due to
statistical error.  This is clear, for example, for the $5^{th}$ and
$8^{th}$ excitations, which have an error much smaller than several
excitations calculated earlier (e.g., $3^{rd}$ and $4^{th}$). The
error in the $3^{rd}$ and $4^{th}$ excitations is mainly due to mixing
among themselves. This result is important because it means that the
present method can be used to calculate several higher excitations
in spite of the errors in lower energy ones.

\begin{figure}
\includegraphics[width=1.\linewidth,clip=true]{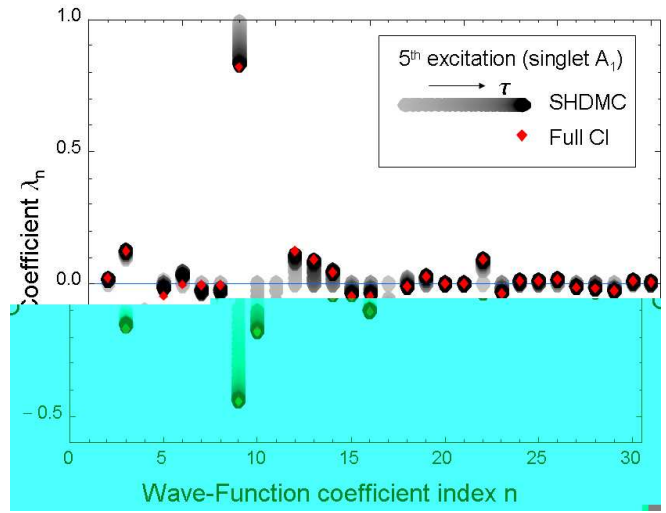}
\caption{(Color online) Change in the values of the multi-determinant
  expansion as the DMC self-healing algorithm progresses for the
  $5^{th}$ excited state of the singlet state of $A_1$ symmetry. Light gray
  colors denote older coefficients, whereas darker ones denote more
  converged results. The full CI results are highlighted
  in small red diamonds.\label{fg:dmcsltex}}
\end{figure}

Figure ~\ref{fg:dmcsltex} shows the evolution of the values of the
coefficients $\lambda_n^{\ell}$ of $|\Psi_T^{\ell}\rangle$ as a function of
the coefficient index $n$ for the $5^{th}$ excited state corresponding to
the singlet configuration of the $A_1$ representation of the group
$D_4$. The shade of gray is light for the older (small $\ell$)
coefficients and deepens to black for the final results (large $\ell$). 
The calculation started from a trial wave-function
orthogonal to the states calculated before as described above.

The coefficients of the wave-function sampled with SHDMC overlap with
the ones obtained with full CI (see Table \ref{tb:overlap}). Similar
results are obtained for all the other excited states calculated.  An
important observation is that the coefficients $\lambda_n$ evolve
continuously towards the exact solution, which suggests the possibility
of accelerated algorithms that extrapolate the values of $\delta
\lambda_n$.

Some eigenstates are significantly more difficult to calculate than
others.  This is typically the case for eigenstates with similar
eigenvalues (e.g., the $6^{th}$ excitation in the singlet case). A
bigger challenge, however, is when $E_L({\bf R})$ is ill behaved,
for example, the case of the $2^{nd}$, $4^{th}$, and $6^{th}$
excitations of the triplet state. Even the full CI wave-function with
300 basis functions has a large variance for $E_L({\bf R})$. In that
case the coefficients obtained with SHDMC and CI are different. This
is due to the fact that the two methods minimize different things: CI
minimizes $\langle~\Psi_n~|(\hat \mathcal{ H}-E_n)^2|\Psi_n\rangle$ on
a truncated basis, and SHDMC minimizes $\int E_L({\bf R}) f({\bf R},\tau) {\bf dR}$
with $\langle \Psi_T | \hat P | \Psi_T \rangle = \langle \Psi_T | \Psi_T \rangle  $.  
Accordingly, the fact that the results are different indicates that
neither calculation, CI or SHDMC, is converged with the basis chosen.
The $4^{th}$ and $6^{th}$ excitations with E symmetry in the triplet
case obtained with SHDMC are a linear combination of the corresponding
ones in full CI.

%start add v6 FAR
\begin{figure}
\includegraphics[width=1.\linewidth,clip=true]{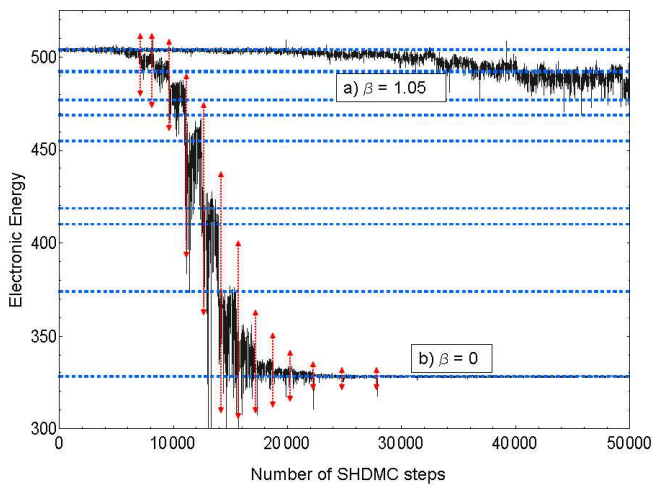}
\caption{\label{fg:drop} (Color online)  Average of the local energy $E_L({\bf R})$ as a 
  function to the number of DMC time steps for two SHDMC runs with
  $\hat P= 1$ starting from a converged trial wave-function
  corresponding to the $8^{th}$ singlet excitation of $A_1$ symmetry
  with a) $\beta = 1.05 $ and b) $\beta =0 $ in Eq.
  (\ref{eq:sampnew}).  The dotted lines  mark the
  beginning of some of the fixed-node DMC blocks of a SHDMC run for
  the $\beta =0 $ case.  Same conventions as in Fig.
  \ref{fg:dmcCIrun:sglex}.}
\end{figure}
Figure \ref{fg:drop} shows the effect of $\hat P$ and $\beta$ [see Eq.
(\ref{eq:sampnew})] on a SHDMC run. The figure shows the average 
of the local energy $E_L({\bf R})$
for two calculations that start from the final trial wave-function
obtained for the $8^{th}$ singlet excitation with $A_1$ symmetry
(please compare it with Fig. \ref{fg:dmcCIrun:sglex}). Both calculations were 
run with the same parameters as in Fig. \ref{fg:dmcCIrun:sglex} 
with two exceptions: i) $\hat P= 1$ was used, which removes 
the orthogonality constraints, and ii) one calculation was run with 
$\beta = 1.05 $ and the other with $\beta=0$ in  Eq. (\ref{eq:sampnew}).
An initial number of blocks $M=20$ was used. 

Both calculations depart from the initial configuration. However, the
run with $\beta = 0$ falls very quickly to the singlet ground state.
The calculation with $\beta=1.05$ remains much longer in the vicinity
of the $8^{th}$ excitation. This clearly shows the stabilizing effect
unequal energy references on excited states.  Since presumably the
$8^{th}$ excitation is not the minimum of its nodal topology, it
finally drifts away. For the $\beta=1.05$ case with $\delta\tau =0.0002$, 
the algorithm becomes
numerically unstable to noise after the $\sim\!50],000$ time step because the variance in
the distribution of weights of the walkers increases and the
statistics is dominated by a reduced number of walkers. 

In contrast,
the first excitation does not drift with $\beta \simeq 1 $ and $\hat P = 1$
(not shown).
 
%% end add FAR

\subsection{Coulomb interaction results and discussion}
\begin{figure}
\includegraphics[width=1.\linewidth,clip=true]{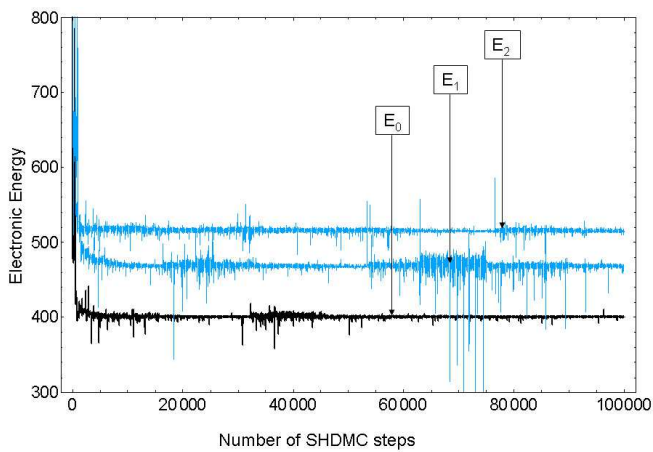}
\caption{\label{fg:coulomb} (Color online) Average of the local energy $E_L({\bf R})$ of
  200 walkers as the SHDMC algorithm converges to the ground, first and
  second eigenstates with $A_1$ symmetry and S=0 of two electrons with
  Coulomb interactions in a square box.  }
\end{figure}

The use of a simplified electron-electron interaction facilitates the
CI calculations and the validation of the optimization method. However,
it is also important to test the convergence and stability of the
method with a realistic Coulomb interaction as in the case of the
ground state.~\cite{keystone} 

The results shown in this section have an interaction potential of the
form\cite{units} $V({\bf r},{\bf r^{\prime}}) = 20 \pi^2 /|{\bf
  r-r\prime}|$ as in Ref.  \onlinecite{keystone}. To mimic the
difficulties that the algorithm would have to overcome in larger or
more realistic systems, the Jastrow term is {\it not} included, i.e.
$J=0$. Most SHDMC parameters are the same as in the model interaction
case.  All calculations with Coulomb interactions were run with
$\beta=0$, the initial number of sub-blocks $M=6$, and the time step
reduced to $\delta \tau = 0.0001$. The initial trial wave-functions
were selected with the criteria used for the model case.

Figure \ref{fg:coulomb} shows the average of the local energy
$E_L({\bf R})$ obtained for the ground state and the
first two excitations with the same symmetry 
(singlet $A_1$). The results are qualitatively similar to those obtained
with the model potential. It is evident from the data that the
variance of $E_L({\bf R})$ and its average are reduced as the
wave-function is optimized. Occasionally, $E_L({\bf R})$ might rise when
$\hat P$ is updated (improving the description of lower energy states).

The energy of the singlet ground state is 400.749 $\pm$ 0.013,  which is
only slightly smaller than the lowest triplet energy\cite{keystone} 
402.718 $\pm$ 0.008 with symmetry $E$.  These energies are very close because of
the dominance of the Coulomb repulsion as compared to the kinetic
energy, which forces the particles to be well separated and therefore
the cost of a node in the triplet state is small. This result is
consistent with the choice of parameters that sets the system 
in the highly correlated regime. 
% begin add FAR
The energies obtained for the first and second excitations 
are\cite{units} $468.56 \pm 0.09$ and $515.50 \pm 0.08$ respectively.

\begin{figure}
\includegraphics[width=1.00\linewidth,clip=true]{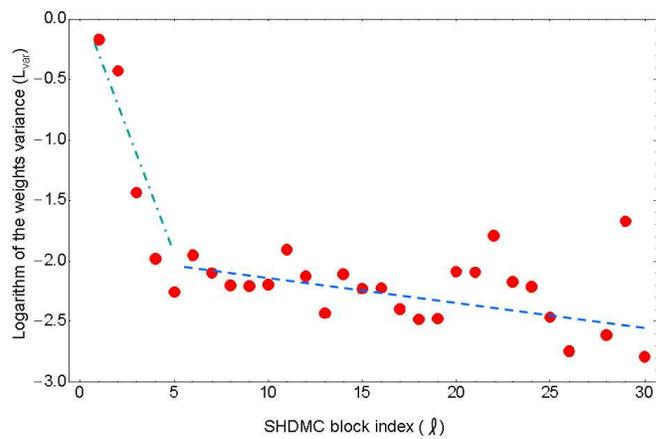}
\caption{(Color online) Logarithm of the variance of the weights of the walkers
  distribution as a function of the SHDMC block index $\ell$ for the
  $2^{nd}$ excitation with $A^1$ symmetry with Coulomb interaction (see Fig.
  \ref{fg:coulomb}). The lines are visual guides.
\label{fg:variance}}
\end{figure}

While Figs. \ref{fg:dmcCIrun:sglex} and \ref{fg:coulomb} are 
qualitatively similar,  the results shown in 
Fig. \ref{fg:dmcCIrun:sglex} are more convincing since they are 
directly compared with full CI calculations and they are less noisy, 
as noted by one referee. 
When the model interaction potential 
is replaced by a Coulomb interaction, full CI calculations are 
still possible, but they involve the numerical calculation of $16471$ 
integrals with Coulomb singularities. 
%This is cumbersome in 
%the context of the symbolic program Mathematica where these calculation
%have been done. 
CI calculations are typically done using a Gaussian
basis,~\cite{dupuis} which limits the impact of the matrix element
integrals of these singularities.  However, as the size of
the system increases, CI calculations become too expansible
numerically. Accordingly, self-reliant methods to validate the quality
of the SHDMC wave-functions must be developed.

As noted earlier, in a fixed population scheme, the weights contain all
the difference between $f({\bf R},\tau)$ and $|\Psi_T({\bf R})|^2$ .
Since $f({\bf R},\tau)$ and $|\Psi_T({\bf R})|^2$ should be equal if
$\Psi_T({\bf R})$ is an eigenstate, the variance of the weights can
be used to measure the quality of the wave-function.  Figure
\ref{fg:variance} shows the evolution of the logarithm of the variance
$L_{var}$ of the weights of the walkers $W_i^j(k)$ [see Eq.
(\ref{eq:weights})] as a function of the SHDMC block index $\ell$.
$L_{var}$ is evaluated as
\begin{equation}
\label{eq:logvar}
L_{var} = \log{\sqrt{\frac{1}{N_c}\sum_{i,j} 
(W_i^{k j}(k) -1)^2 }} \; . 
\end{equation}
By using a linear order expansion in $\delta\tau$ in Eq.
(\ref{eq:weights}) and using Eq. (\ref{eq:enaverage}), it is straightforward to relate Eq.
(\ref{eq:logvar}) to the variance of $ E_i^j(k)$. The latter is an average of $E_L({\bf R})$.
A common measure of the quality of the ground state wave-function is the variance of
$E_L({\bf R})$.   

The results shown in Fig. \ref{fg:variance} correspond to the $2^{nd}$
singlet excitation with $A_1$ symmetry (see Fig. \ref{fg:coulomb}).
Similar results are obtained for the ground state and the first
excitation (not shown).  The error bar in $L_{var}$ is smaller than
the size of the symbols. The fluctuations in $L_{var}$ result from
the random fluctuations of the coefficients $\lambda_n$ that are
obtained statistically. Note that in spite of the noise, a clear
trend shows the improvement of the quality of the
wave-function and $E_T$ as the SHDMC algorithm progresses. However, these
improvements are not uniform, which is
reflected by the oscillations in $L_{var}$ in Fig. \ref{fg:variance}
and in the amplitude of $E_L({\bf R})$ in Fig. \ref{fg:coulomb}. A
careful user of SHDMC should track $L_{var}$ and use the best quality
wave-function to calculate energies and $\hat P$.

% end add FAR
\section{Summary}
\label{sc:discussion}
An algorithm to obtain the approximate nodes, wave-functions, and
energies of arbitrary low-energy eigenstates of many-body Hamiltonians
has been presented. This algorithm is a generalization of the ``simple''
self-healing diffusion Monte Carlo method developed for the
calculation of the ground state of fermionic systems,\cite{keystone}
which in turn is built upon the standard DMC method.~\cite{ceperley80}

At least in the case of the tested system, wave-functions and energies
that continuously approach fully converged configuration interaction
calculations can be obtained depending only on the computational time.
The wave-function, in turn, allows the calculation of any observable. 

It is found that some special eigenstates, presumably 
the minimum energy eigenstate for a given nodal topology, can be
obtained without calculating the lower excitations by artificially
generating a kink in the propagated function using unequal energy
references in different nodal pockets.

The present method can be implemented easily in existing codes.
Ongoing tests on the ground state method\cite{keystone} in larger
systems give serious hope\cite{fn:tests} that the current
generalization will also be useful.

While there are methods to obtain the excitation spectra of a
many-body Hamiltonian in a variational Monte Carlo context~\cite{kent98,umrigar07} they require
obtaining the Hamiltonian and the overlap matrix elements.
This requirement would present a challenge for very large systems.
SHDMC is a complementary technique that could potentially scale better
for larger sizes.  The evaluation and storage of the matrix elements
of $\hat \mathcal{H}$ is not required. The number of quantities
sampled [the projectors $\xi_n ({\bf R})$, Eq. (\ref{eq:xi})] is equal
to the number of basis functions $n_b$. In contrast, energy
minimization methods or configuration interaction (CI) require the
evaluation of $n_b^2$ matrix elements. In addition, the solution of a
generalized eigenvalue problem with statistical noise is avoided. This
can be an advantage in very large systems since algorithms for
eigenvalue problems are difficult to scale to take maximum advantage
of large supercomputers.  In contrast, the sampling of a large number
of determinants can be trivially distributed on different processors.
Moreover, recent advances in determinant evaluation could facilitate
sampling a very large number of projectors $\xi_n ({\bf
  R})$.~\cite{nukala09}

An apparent disadvantage of SHDMC is that the method is recursive.
This disadvantage is partially removed since i) the number of blocks
$M$ used to collect data is increased only if necessary to
improve the wave-function significantly,~\cite{fn:changes} ii) and, the
propagation to large imaginary times is avoided by using precisely this
recursive approach that accumulates the propagation in successive
blocks. In addition, a small value of $k \; \delta \tau $ limits large
fluctuations in the weights, which recently have been claimed to cause
an exponential cost in the convergence of DMC results.~\cite{nemec09}

The dominant cost of the present algorithm to obtain the
wave-functions and their nodes scales as $N_e^3 \times n_{max} \times
n_b \times n_{st}$, with $n_{max}$ being the number of excited states,
$n_b$ the number of projectors $\xi_n({\bf R})$ sampled, and $n_{st}$
the total number of SHDMC steps.  Of course, the error and the cost
depend on the quality of the method used to construct $\Phi_n({\bf
  R})$ and the quality of the initial trial wave-functions. Systematic
errors decrease when $n_b$ is large, and the statistical error
decreases when $n_{st}$ increases.  For a fixed absolute error, $n_b$
is expected to increase exponentially with the number of electrons
$N_e$.~\cite{keystone}

Note that in order to describe an arbitrary wave-function of a system
with $N_e$ electrons and a typical size $L$ in $D>1$ dimensions with a
resolution $R_s$, one needs approximately $(L/R_s)^{(D\; N_e)}$ basis
functions.  The nodal surface alone requires ~$(L/R_s)^{(D\; N_e-1)}$
degrees of freedom. Therefore, finding an algorithm to obtain the
nodes $S_n({\bf R})$ of any eigenstate $n$ with an arbitrary
interaction in a time polynomial in $N_e$ is potentially a
``Philosopher's Stone'' quest.  However, if exponential factors
actually control the accuracy of the DMC approach, as
claimed,~\cite{nemec09} just a rock solid method to find the nodes
which simultaneously improves the wave-function (reducing the
population fluctuations) could be considered a satisfactory solution.
The presented work could be the basis of such a method.

In ongoing work, SHDMC methods are being developed and tested in
larger systems.

\section*{Acknowledgments}
  The author would like thank C.
Umrigar for suggesting the sampling of $\delta \lambda_n$ instead of
the absolute value of the coefficients.
The author also thanks R. Q. Hood, M. Bajdich and P. R. C. Kent for
a critical reading of the manuscript and for related discussions. Finally,
the author thanks the anonymous referee who inspired the calculations
presented in Figs. \ref{fg:drop} and \ref{fg:variance}.

Research performed at the Materials Science and Technology Division
sponsored by the Department of Energy and the Laboratory Directed
Research and Development Program of Oak Ridge National Laboratory,
managed by UT-Battelle, LLC, for the U.S. Department of Energy under
Contract No. DE-AC05-00OR22725.


\begin{thebibliography}{99}
\bibitem{fn:highly} Understanding sections I, II, III, V and VI of Ref.
  \onlinecite{keystone} is required before reading this article.
\bibitem{fn:nodelta} Note that the limit $\epsilon \rightarrow 0$ is taken
after the amplitude of the potential tends to $\infty$. Thus, this potential does not
have the $\delta({\bf R})$ form, and every eigenstate of $\mathcal{H}_{FN} $
must be zero at $S_T({\bf R})$.
\bibitem{fn:Jastrow} The Jastrow factor does not change the nodes but accelerates
convergence and improves the algorithm's numerical stability.
\bibitem{fn:groundstate} The ground state $|\Psi_0\rangle $ is formally obtained 
byapplying the evolution
operator $e^{-\tau \hat\mathcal{H}}$ to a trial wave-function 
$|\Psi_T^{\ell=0}\rangle $ in the limit $\tau~\rightarrow~\infty $.
\bibitem{fn:permutations} All symmetries of $\hat \mathcal{H}$ must be
considered, which includes space group symmetries, spin, and particle
permutations.
\bibitem{fn:changes} If the change in the wave-function coefficients
  is dominated by random noise, the scalar product between
  the old and the new $\delta\lambda_n$ can be negative and $M$ is
  multiplied by a factor larger than 1.
\bibitem{fn:tests} A SrLi dimer with $13$ electrons has been compared
  with energy minimization calculations.\cite{umrigar07} We have also
  a proof of principle for C$_{20}^{+2}$ (78 electrons and ~700
  determinants).
\end{thebibliography}
\end{document}